\documentclass{elsart}
\usepackage{graphicx}

\usepackage{amssymb}

\begin{document}

\begin{frontmatter}
\title{Electron-phonon induced spin relaxation in InAs quantum dots}
\author[ufscar]{A. M. Alcalde\thanksref{email}},
\author[ufu]{Qu Fanyao},
\author[ufscar]{G. E. Marques}

\address[ufscar]{Departamento de F\'{\i}sica, Universidade Federal de S\~ao Carlos, 13565-905, S\~ao Carlos, SP, Brazil}
\address[ufu]{LNMIS, Faculdade de F\'{\i}sica, Universidade Federal de Uberl\^andia, 384000-902 Uberl\^andia, MG, Brazil}
\thanks[email]{E-mail: palcalde@df.ufscar.br, Phone: +55 16 2608205,
Fax: +55 16 2614835}

\begin{abstract}
We have calculated spin relaxation rates in parabolic quantum dots due to the
phonon modulation of the spin-orbit interaction in presence of an external magnetic field.
Both, deformation potential and piezoelectric electron-phonon coupling mechanisms are included within the
Pavlov-Firsov spin-phonon Hamiltonian. Our results have demonstrated that, in narrow gap materials,
the electron-phonon deformation potential and piezoelectric coupling give comparable
contributions as spin relaxation processes.
For large dots, the deformation potential interaction becomes dominant.
This behavior is not observed in wide or intermediate gap semiconductors, where
the piezoelectric coupling, in general, governs the spin relaxation processes.
We also have demonstrated that spin relaxation rates are particularly sensitive to the Land\'e $g$-factor.

\end{abstract}

\end{frontmatter}


The ability to manipulate and control processes that involve transitions
between spin states is, at the moment, of extreme importance due to the
recent applications in quantum computation and quantum communication.
Quantum dots (QD's) of diverse geometries are good candidates for
implementation of semiconductor quantum communication devices because the
electronic, magnetic and optical properties can be controlled in the modern
grown and nanofabrication techniques. The spin dephasing time is important
because it sets the length-scale on which coherent physics can be observed.
It is therefore important to understand the origin of decoherence so that
ultimately it may be reduced or controlled. At the moment remains in
discussion which, between these processes, is dominant in semiconductors in
low dimensional systems. Some experimental results have shown good agreement
with the theoretical predictions in 2-D systems~\cite{Lau01} but, in
general, the identification of the processes through direct comparison with
the experimental results can become a formidable task. This problem is more
critical in QD's, since few experimental results exist and the theoretical
discussion of the spin relaxation mechanisms is still an open subject.
Khaetskii and Nazarov~\cite{Khaetskii01} studied two kind of processes of
spin relaxation induced by phonons in GaAs QD's: i) The first mechanism is
due to the Dresselhaus admixture of states with an opposite spin. Without
the spin-orbit interaction the Zeeman sublevels correspond to the orbital
state with ``pure" spin states. The addition of spin-orbit
terms provides a small admixture of the states with opposite spin to each
sublevel and, thus, enables the phonon-assistant transition between them;
ii) The second process is related with the direct coupling between the spin
and the strain field as produced by the acoustic phonons. With admixture
process (i), scattering rates of the order of $10^{3}$ s$^{-1}$ were
obtained. The authors also showed that the admixture process is the dominant
one. Recently, Woods \textit{et al,}~\cite{Woods02} have studied spin
relaxation considering the coupling between spins and phonons arising from
interface motion. This mechanism arises when interface motion due to
acoustic phonons changes a parameter of the system. In this case, small ($%
10^{3}$ s$^{-1}$) and strongly size dependent relaxation rates have been
found by the authors. In these works, the spatial dependence of the $g$-factor
was not considered.
Also, due to the small transition energies,
only the coupling by piezo-phonons were included in the rate calculation.
These approximations are not necessarily valid for InAs QD's, since: i)
Experimental measurements have shown that the electron $g$-factor depends
strongly on the dot size~\cite{Thornton98,Medeiros02}, ii) In InAs based
QD's, the Zeeman transition energies can be considerably larger than for GaAs
QD's and, therefore, the coupling due to the deformation potential can
contribute significantly to the relaxation rates.

Our approach is based on the model of Pavlov and Firsov\cite%
{Pavlov66,Pavlov67}. In this model, the Hamiltonian describing the
transitions with spin reversal, in the scattering of electrons by phonons,
can be written in a general form, $H_{e-ph}^{\sigma }=U_{ph}+\beta \left[
\mathbf{\sigma }\times \nabla U_{ph}\right] \cdot \left( \mathbf{p}+\frac{e}{%
c}\mathbf{A}\right) $, where $U_{ph}$ is the phonon operator, $(\hbar /2)%
\mathbf{\sigma }$ is the spin operator, $\mathbf{p}$ is the momentum
operator and $\mathbf{A}$ is the vector potential for the magnetic field $%
\mathbf{B}$. This interaction Hamiltonian depends on spin variables and,
thus, can lead to spin-flip transitions between pure spin states. In this
work we have calculated spin relaxation rates in InAs parabolic quantum
dots, considering the phonon modulation of the spin-orbit interaction within
the Pavlov-Firsov model. We also study the effects of the spatial dependence
of the electron $g$-factor and evaluate the contributions of the deformation
potential and piezoelectric couplings on the spin relaxation rates.


Experimental measurements and numerical calculations~\cite{Hawrylak99} have
indicated that in lens-shaped quasi-two dimensional self-assembled quantum
dots the bound states of both electrons and holes can be understood assuming
an effective parabolic potential, $V(\rho )=\frac{1}{2}m\omega _{0}^{2}\rho
^{2}$, where $\hbar \omega _{0}$ is the characteristic confinement energy
and $\rho $ is the radial cylindrical coordinate. By using a one-band
effective mass approximation and considering the presence of a magnetic
field $B$, applied normal to plane of the dot, one can write the electron
wavefunctions as~\cite{Voskoboynikov01}
\begin{equation}
f_{n,l,\sigma } = \left[{\frac{{n!}}{{\pi (n+|l|)!}}}\right] ^{\frac{1}{2}}%
\frac{\rho ^{|l|}}{a^{|l|+1}}e^{-\frac{\rho ^{2}}{2a^{2}}}e^{il\varphi}
L_{n}^{|l|}\left( {\frac{\rho ^{2}}{a^{2}}}\right)
\chi(\sigma).
\label{funciondeonda}
\end{equation}%
In the above expression $L_{n}^{|l|}$ denotes the Laguerre polynomials, $n$
is the principal quantum number, $l$ is the azimuthal quantum number and $%
\chi (\sigma )$ is the spin wavefunction for spin variable $\sigma $. The
corresponding eigen-energies are $E_{n,l,\sigma }=(2n+|l|+1)\hbar \Omega
+(l/2)\hbar \omega _{c}+(\sigma /2)g\mu _{B}B,$ where $\Omega =(\omega
_{0}^{2}+\omega _{c}^{2}/4)^{1/2}$, $\mu _{B}$ is the Bohr magneton, $%
a=(\hbar /m\Omega )^{1/2}$ is the effective length and $\omega _{c}=eB/m$.
The Land\'{e} $g$-factor and the effective mass $m$ are expressed in
second-order $\mathbf{k}\cdot \mathbf{p}$ perturbation.~\cite{Roth59}
\begin{equation}
g=2-\frac{4m_{0}P^{2}}{3\hbar ^{2}}\frac{\Delta }{\left( E_{g}+E\right) %
\left[ \left( E_{g}+E\right) +\Delta \right] },  \label{factorg}
\end{equation}%
\begin{equation}
\frac{1}{m}=\frac{1}{m_{0}}+\frac{2P^{2}}{3\hbar ^{2}}\frac{3\left(
E_{g}+E\right) +2\Delta }{\left( E_{g}+E\right) \left[ \left( E_{g}+E\right)
+\Delta \right] }.  \label{masa}
\end{equation}%
Here, $\Delta $ is the spin-orbit splitting, $E_{g}$ is the energy band gap,
$E$ is the electron energy measured from the bottom of the conduction band
and $P=(\hbar /m_{0})\langle iS|p_{z}|Z\rangle $ represent the interband
matrix element. For InAs the value of $\Delta $ is comparable with the
fundamental gap $E_g$, thus we can expect significant variations of the electron $%
g $-factor with the size parameters.

For an assisted acoustic-phonon spin-flip process, the matrix element, $M,$
for electron spin-flip between initial ( $|nl\uparrow \rangle $ ) and final
( $|n^{\prime }l^{\prime }\downarrow \rangle $) states with emission of a
phonon of momentum $\mathbf{q}$ and energy $\hbar vq$, can be obtained, from
the Pavlov-Firsov spin-phonon Hamiltonian~\cite{Pavlov66,Pavlov67} as
\begin{eqnarray}
M_{n l \uparrow \rightarrow n^\prime l^\prime \downarrow}&=& d(q)\left( \frac{\hbar}{\rho_M V v q}\right)^{1/2} \nonumber \\
& &\chi_z(\uparrow)
\left(\begin{array}{cc}
0 &\mathbf{\hat{n}}^-\times\mathbf{\hat{e}_q}  \\
\mathbf{\hat{n}}^+\times\mathbf{\hat{e}_q} & 0
\end{array}
\right) \chi_z(\downarrow) \cdot \nonumber \\
& & \int d^3{\mathbf r} f_{n^\prime l^\prime}e^{-i\mathbf{q}\cdot \mathbf{r}}
\left( \frac{\mathbf{p}}{\hbar} + \frac{e\mathbf{A}}{\hbar c} + \mathbf{q} \right)
f_{nl}, \nonumber \\
\label{spinphonon}
\end{eqnarray}
where $\chi _{z}$ are the spin wavefunctions quantized along the $z$ axis, $%
f_{nl}=\langle \mathbf{r}|n,l\rangle $ is the electron envelope
wavefunction, the magnetic vector potential $\mathbf{A}$ is obtained in the
symmetric gauge considering that the orientation of $\mathbf{B}$ coincide
with the $z$-axis. $\mathbf{\hat{n}}^{\pm }=\mathbf{\hat{x}}\pm i\mathbf{%
\hat{y}}$, 
$\mathbf{\hat{e}_{q}}=\mathbf{q}/q$ is the polarization vector of the
longitudinal acoustic phonons, $v$ is the average sound velocity, $\rho _{M}$
is the mass density, $V$ is the system volume and $d$ is a coupling constant
that depends on the electron-acoustic phonon coupling mechanism. Detailed
expressions for the parameter $d$ can be found in Ref.~\cite{Pavlov67}.

The spin-flip transition rate $W$ is calculated from the Fermi Golden Rule
\begin{equation}
W=\frac{{2\pi }}{\hbar }\frac{V}{{(2\pi )^{3}}}\int {d^{3}\mathbf{q}}%
\left\vert M_{nl\uparrow \rightarrow n^{\prime }l^{\prime }\downarrow
}\right\vert ^{2}\delta (\hbar vq-\Delta E),  \label{fermi}
\end{equation}%
where $\Delta E=E_{nl\uparrow }-E_{n^{\prime }l^{\prime }\downarrow }$ is
the transition energy.


The calculations were performed at $T\sim 0$~K and we only have considered
transitions between ground state Zeeman levels. The temperature dependence
for one-phonon emission rate is determined from $W=W_{0}(n_{B}+1)$, where $%
n_{B}$ is the Bose distribution function and $W_{0}$ is the rate at $T=0$~K.
In the temperature regime $T\lesssim $10~K and considering typical values of
magnetic field $(B\sim 2~\mathrm{T})$, the Bose function is $n_{B}+1\approx
1 $ and $W\approx W_{0}$. For temperatures larger than few Kelvin degrees,
two-phonon processes should be considered as the dominant spin relaxation
mechanism. These type of processes have not been considered in the present
calculation.

In general, we obtain that $W\sim (g\mu _{B}B)^{k}$, $k$ being an integer
number that depends on the electron-phonon coupling process ($k$ = 7 for
deformation potential and $k$ = 5 for piezoelectric coupling). This strong
dependence with the transition energy and, in consequence with the $g$%
-factor, demands that this parameter should be determined taking in account
the effects of the quantum confinement. Spin splitting measurements in InAs
self-assembled QD's~\cite{Thornton98,Medeiros02} have revealed $g$-factors
showing clear dependence on the dot size. Values in the range of 0.8 - 1.29 were
reported for QD's in strong confinement regime and they differ
strongly from the value of bulk InAs ($g_{bulk}$= - 14.4). In order to
include the spatial dependence of the $g$-factor in the rate calculation we
have used the Roth formula given in Eq. (\ref{factorg}).

\begin{figure}[tbp]
\begin{center}
\includegraphics*[scale=1.0]{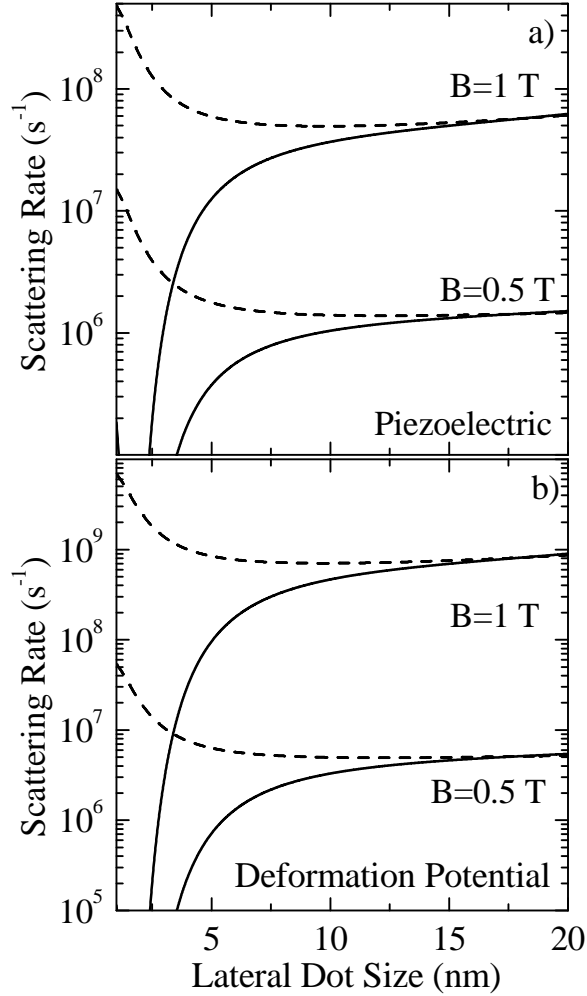}
\end{center}
\caption{Spin-relaxation rates, for InAs QD's, as a function of the lateral
dot size $R$ : a) piezoelectric and b) deformation potential mechanisms.
Rates calculated from Eq.~(\protect\ref{factorg}) ( for $g=g_{bulk}$) are
shown in solid (dashed) lines.}
\label{figure1}
\end{figure}

These facts are clearly illustrated in Fig.~\ref{figure1}, where we show
calculated InAs QD spin relaxation rates, for piezoelectric and deformation
potential couplings, as a function of the lateral dot size $R$ . Rates with $%
g=g_{bulk}$ are shown in dashed lines, and those with $g$ given by Eq.(\ref%
{factorg}), in solid lines.

Our results show that the effect of confinement on the $g$-factor produces
significant variations in the relaxation rates. We observe that the rates
obtained for $g=g_{bulk}$ are one order of magnitude larger than rates
calculated from Eq.~(\ref{factorg}) for deformation potential mechanism, $B$
= 1~T and $R$ = 5~nm.

The behavior of the rates with the QD size depends directly on the spatial
overlap integral and the magnitude and on the sign of the $g$-factor. For
negative values of $g$, the rate diminishes as the dot size is increased.
This behavior can be illustrated in the rates calculated with $%
g=g_{bulk}=-14.4$ (dashed lines in Fig.~\ref{figure1}). The Eq.(\ref%
{factorg}) provides positive values of $g$ for $R\lesssim 10$ nm, in this
case the rate increases as the dot size increases (solid lines in Fig.~\ref%
{figure1}). The $g$-factor dependence on size can be neglected in the
relaxation rates, for dots with $R>15$~nm, since the energy of Zeeman level
becomes very small.

\begin{figure}[tbp]
\begin{center}
\includegraphics*[scale=1.0]{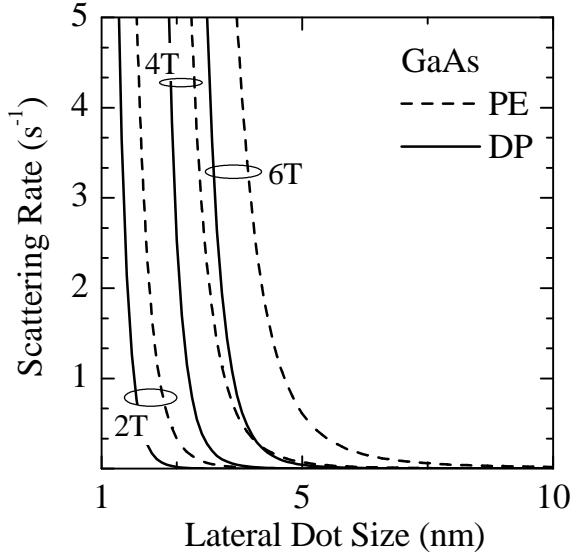}
\end{center}
\caption{Spin-relaxation rates for an GaAs QD as a function of the lateral dot size $R$ for
piezoelectric (dashed lines) and deformation potential (solid lines) mechanisms.}
\label{figure2}
\end{figure}

According with the general relation, $W \sim (g\mu _{B}B)^{k}$, we can also
observe that the rates will depend strongly on the magnetic field and can
increases several order of magnitude when values of $B$ are swept from 0.5~T
to $1$~T. Our results demonstrate that for narrow-gap materials, the
piezoelectric [Fig.~\ref{figure1}a)] and deformation potential [Fig.~\ref{figure1}b)]
coupling present comparable contribution to the spin relaxation process.
Indeed, for large dots and large magnetic fields, the deformation potential
interaction becomes the dominant one. This result is not observed in GaAs
(see Fig.~\ref{figure2}), where the PE coupling, in general, governs the
relaxation process.

For GaAs case, the confinement does not produce important modifications in
the $g$-factor and the scattering rates shown in the Fig.~\ref{figure2}
reveal the negative character of the $g$-factor for the considered dot
sizes.
Also, the Fig.~\ref{figure2} shows that the GaAs rates diminish really fast
as the dot size increases. For $R>10$~nm, the relaxation times can be as
large as seconds and the rates become almost independent of $R$. This
behavior as well as the magnitude of rates are similar to those obtained by
Woods~\cite{Woods02} when considered other relaxation mechanisms mediated by
acoustic phonons.


In conclusion, we have studied the spin relaxation of electrons in InAs and
GaAs parabolic quantum dots by considering the phonon modulation induced on
the spin-orbit interaction as the relaxation process. For dots based on
narrow-gap materials, we have found that the size dependence (spatial
localization) of the $g$-factor cannot be neglected in the calculation.
Also, the deformation potential mechanism can become dominant, especially
for large $B$.

\section*{Acknowledgments}

This work has been supported by Funda\c{c}\~{a}o de Amparo \`{a} Pesquisa do
Estado de S\~{a}o Paulo (FAPESP) and by Conselho Nacional de Desenvolvimento
Cient\'{\i}fico e Tecnol\'{o}gico (CNPq).

\end{document}